# $Re^{4+}$ Luminescence as a Highly Sensitive Alternative to Ruby for Optical Pressure Sensing

**Yeshan Wu[a†], Maja Szymczak[b†], Lukasz Marciniak[b*], Yifei Yu[a], Shuailing Ma[a*], Tian Cui[a], Laihui Luo[a*], Peng Du[a*]**

[a]*School of Physical Science and Technology, Ningbo University, 315211 Ningbo, Zhejiang, China*

[b]*Institute of Low Temperature and Structure Research, Polish Academy of Sciences, Okólna 2, 50-422 Wrocław, Poland*

**Abstract**

Luminescence-based remote pressure sensing provides a powerful approach for pressure determination under conditions where conventional contact methods are difficult to implement. Although ruby remains the undisputed gold standard among luminescent pressure indicators, its relatively low pressure sensitivity and susceptibility to temperature variations represent important limitations. Therefore, in this work, $Cs_2HfCl_6:Re^{4+}$ is proposed as an alternative luminescent pressure indicator exhibiting a 15-fold higher pressure sensitivity (6.93 nm $GPa^{-1}$) than ruby while maintaining a comparable thermal sensitivity, making it a particularly attractive material for optical manometry. Furthermore, $Cs_2HfCl_6:Re^{4+}$ enables ratiometric pressure readout, providing a high relative pressure sensitivity reaching 175.1% $GPa^{-1}$. To the best of our knowledge, this work represents the first demonstration of $Re^{4+}$ luminescence for pressure sensing, introducing a new class of luminescent pressure indicators and opening new opportunities for the development of highly sensitive optical manometers based on $Re^{4+}$-activated materials.

†Y. Wu and M. Szymczak contributed equally in this work

**Corresponding authors**

***E-mail:** l.marciniak@intibs.pl (L. Marciniak); mashuailing@nbu.edu.cn (S. Ma); luolaihui@nbu.edu.cn (L. Luo); dupeng@nbu.edu.cn (P. Du)

## Introduction

Pressure, alongside temperature, is one of the most fundamental thermodynamic parameters and plays a crucial role in determining numerous physical and chemical phenomena, particularly the properties of materials allowing *in-situ* manipulating the intrinsic properties of materials, including bond lengths, electronic structures, and crystal structures, thereby enabling the discovery of unconventional phenomena, such as pressure-induced luminescence modulation, superconductivity under extreme conditions, and the synthesis of novel materials [1–3]. From this perspective, understanding the stability of materials and the evolution of their electrical, mechanical, optical, structural, and magnetic properties under high-pressure conditions is of considerable importance. Such investigations provide invaluable insights into structure-property relationships and enable access to material states and phenomena that are inaccessible under ambient conditions[4–6]. Consequently, high-pressure studies have been extensively conducted using various types of pressure cells, among which the diamond anvil cell (DAC) has become one of the most widely employed[7–10]. The DAC enables the generation of extremely high pressures while maintaining optical access to the investigated sample through the diamond anvils[11]. This unique feature allows a broad range of in situ spectroscopic and optical measurements to be performed under well-controlled high-pressure conditions.

Reliable determination of the pressure generated inside a DAC requires the use of an appropriate pressure calibrant. One of the most commonly employed materials for this purpose is $Al_2O_3:Cr^{3+}$, commonly known as ruby[6,11–17]. Under ambient conditions, ruby exhibits characteristic narrow emission lines near 690 nm associated with the $^2E \rightarrow {}^4A_2$ electronic transition of $Cr^{3+}$ ions. Application of pressure compresses the crystal lattice and decreases the $Cr^{3+}$-$O^{2-}$ interionic distances, thereby modifying the nephelauxetic effect experienced by the $Cr^{3+}$ ions[6,13,14]. Consequently, increasing pressure induces a monotonic redshift of the $Cr^{3+}$ emission lines. By monitoring their spectral position, the pressure generated inside the DAC can be accurately determined. The narrow spectral character of the ruby emission is particularly advantageous in this respect, as it enables precise determination of the emission line position and, consequently, accurate hydrostatic pressure readout. An additional advantage of ruby as a pressure calibrant is its excellent mechanical and chemical stability, which enables its application for pressure determination up to approximately 150 GPa [11].

Despite these considerable advantages, an important limitation of ruby is the relatively modest pressure-induced spectral shift of its emission lines, which can limit the accuracy and resolution of pressure determination. This issue becomes particularly relevant when considering that the spectral position of the ruby emission is also temperature dependent[18]. Consequently, under experimental conditions in which both pressure and temperature vary, temperature-induced spectral shifts may introduce substantial uncertainty into the pressure readout.

Therefore, considerable research efforts have been devoted to identifying alternative luminescent pressure calibrants that offer enhanced pressure sensitivity while maintaining a more favorable balance between pressure and temperature sensitivities[5,14,19–30]. Numerous materials exploiting the $^2E \rightarrow ^4A_2$ emission of $Cr^{3+}$ ions incorporated into host lattices other than $Al_2O_3$ [5,20,31], as well as systems based on alternative luminescent ions[21–23,26,27,29,30], have consequently been investigated. Changing the luminescent ion, however, typically modifies the spectral position of the pressure-sensitive emission. In certain applications, such a shift may be advantageous, for example by minimizing spectral overlap between the emission of the pressure calibrant and that of the material under investigation. Nevertheless, the red-to-near-infrared spectral region characteristic of the $^2E \rightarrow ^4A_2$ emission band of $Cr^{3+}$ remains particularly attractive for a broad range of high-pressure experiments. The identification of new luminescent materials combining narrowband red/NIR emission with substantially enhanced pressure sensitivity is therefore highly desirable.

In this context, to the best of our knowledge, we report here for the first time a systematic high-pressure investigation of the luminescence properties of $Re^{4+}$ ions in $Cs_2HfCl_6$. The $^2T_{2g}(\Gamma_7) \rightarrow ^4A_{2g}$ electronic transition, giving rise to a narrow emission band centered at approximately 730 nm, exhibits a pressure-induced spectral shift more than 15 times greater than that of ruby within the investigated pressure range up to approximately 7 GPa. Importantly, $Cs_2HfCl_6:Re^{4+}$ simultaneously exhibits a temperature sensitivity comparable to that of ruby, resulting in a substantially more favorable balance between pressure and temperature sensitivities. This feature is particularly advantageous for minimizing temperature-induced uncertainty in pressure determination. Furthermore, we demonstrate that $Cs_2HfCl_6:Re^{4+}$ can be employed as a ratiometric luminescent pressure sensor, exhibiting one of the highest relative pressure sensitivities reported to date. These results identify $Cs_2HfCl_6:Re^{4+}$ as a promising alternative to ruby for highly sensitive and precise optical pressure determination and highlight the potential of $Re^{4+}$-based luminescence for the development of a new class of high-performance pressure calibrants.

## Results

The $Cs_2HfCl_6:Re^{4+}$ investigated in this study can be synthesized using a straightforward and relatively low-cost procedure, as schematically illustrated in Figure 1a. In the structure of this phosphor each Hf atom is coordinated with six Cl atoms, resulting in the formation of an isolated $[HfCl_6]^{2-}$ octahedron, while Cs atom occupies the interstitial sites between these octahedral units[32–41] (Figure 1b). When $Re^{4+}$ is traduced, it is expected to replace the $Hf^{4+}$ site and forms the $[ReCl_6]^{2-}$ octahedron (Figure 1b). The comparison of the XRD patterns of $Cs_2HfCl_6:Re^{4+}$ with different dopant ions concentrations reveals that diffraction profiles remain unchanged as $Re^{4+}$ doping content increases, and the corresponding diffraction peaks are well indexed to the standard cubic $Cs_2HfCl_6$ (PDF#32-0233), confirming the single phase of the studied samples (Figure 1c). In order to get deeper insight into the structural characteristics of the designed compounds, Rietveld refinements of the XRD patterns of the representative double perovskites were conducted, as presented in Figure 1d) and S1. Evidently, these refined profiles coincide well with the measured data, further manifesting the high phase purity and cubic crystal structure of the synthesized compounds. Moreover, owing to the difference in the ionic radii between $Hf^{4+}$ (*i.e.*, 0.71 Å) and $Re^{4+}$ (*i.e.*, 0.63 Å)[42], the lattice contraction is observed at the elevated dopant ions concentration, as confirmed by the decreased lattice parameters (*i.e.*, $a = b = c$) and cell volume (see Table S1). The measurement of the Raman spectrum of the $Cs_2HfCl_6$:$0.008Re^{4+}$ double perovskite reveals the presence of two intense peaks at around 163.7 and 329.6 $cm^{-1}$ pertaining to the $T_{2g}$ and $A_{1g}$ vibrational modes of the $[HfCl_6]^{2-}$ octahedron[43], respectively (Figure 1e). These results suggest that the designed $Cs_2HfCl_6$:$xRe^{4+}$ double perovskites with pure cubic phase are successfully synthesized and the doping of $Re^{4+}$ in the analyzed range does not affect the phase purity of the analyzed phosphor.

The morphological details of the resulting products were examined by SEM and TEM measurements. From the SEM image (Figure 1f), it can be seen that the synthesized double perovskite comprises irregular microparticles, and neither the particle morphology nor size exhibits an obvious dependence on $Re^{4+}$ concentration, as depicted in Figure S2a) and S2b). The high-resolution TEM image presented in Figure 1g contains distinct lattice fringes, with an average distance of around 2.94 Å corresponding to the (222) crystal plane of cubic $Cs_2HfCl_6$ (PDF#32-0233). Moreover, many bright diffraction spots are observed in the selected-area electron diffraction (SAED) pattern (Figure 1h), which implies the polycrystalline characteristics and good crystallinity of the designed compounds. Furthermore, via using the EDS spectrum (Figure S2c), the elements (*i.e.*, Cs, Hf, Cl and Re) presented in the designed double perovskites are explored, and they are found to be uniformly distributed throughout the whole microparticles, as proved by the elemental mapping results (Figure 1i-l).

Through using X-ray photoelectron spectroscopy (XPS) measurement, we further examined the elemental compositions and their valences in the target materials. The full-survey

XPS spectrum clarifies that the studied samples contain the elements of Cs, Hf, Cl and Re (Figure S3a). The high-resolution XPS spectrum of $Cs^{+}$ 3d contains two intense bands at 724.8 and 738.8 eV pertaining to $Cs^{3+}$ $3d_{5/2}$ and $Cs^{3+}$ $3d_{3/2}$, respectively[44,45], as shown in Figure S3b), while that of the $Hf^{4+}$ 4d can be fitted into two peaks with the binding energies of 17.8 and 19.6 eV, which are attributed to the $Hf^{4+}$ $4d_{7/2}$ and $Hf^{4+}$ $4d_{5/2}$, respectively (Figure S3c)[45]. Moreover, the high-resolution XPS spectrum of $Cl^{-}$ 2p presented in Figure S3d) is also able to be divided into two bands at 198.9 and 200.7 eV, which are assigned to $Cl^{-}$ $2p_{3/2}$ and $Cl^{-}$ $2p_{1/2}$, respectively[44,45]. Furthermore, the formation of $Re^{4+}$ in the synthesized double perovskites is verified by its high-resolution XRD spectrum (Figure S3e), in which two peaks at 43.9 and 46.3 eV corresponding to $Re^{4+}$ $4f_{7/2}$ and $Re^{4+}$ $4f_{5/2}$ are observed. These findings indirectly prove the success synthesis of the $Re^{4+}$-doped $Cs_2HfCl_6$ double perovskites.

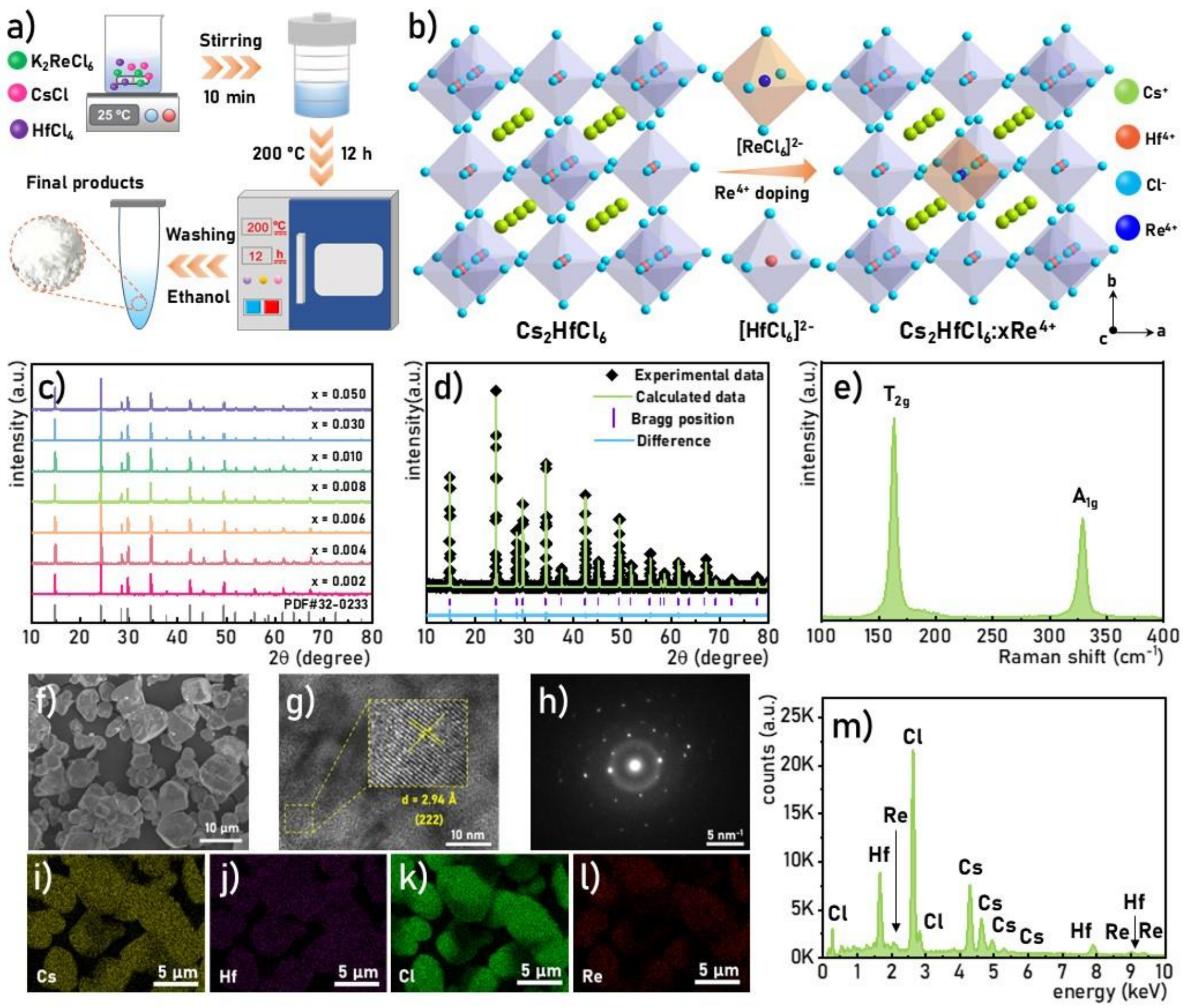

**Figure 1** Schematic illustration of the preparation process of the $Cs_2HfCl_6$: $Re^{4+}$ -a); visualization of the structure of $Cs_2HfCl_6$ and $Cs_2HfCl_6$: $Re^{4+}$ compounds- b); XRD patterns of the $Cs_2HfCl_6$:$x$$Re^{4+}$ with different dopant ions concentrations -c); Rietveld XRD refinement -d), room temperature Raman spectrum -e), SEM image -f), high-resolution TEM image -g), SAED pattern -h) and elemental mapping - i) - l) of the $Cs_2HfCl_6$:0.008$Re^{4+}$; full survey XPS spectrum of the $Cs_2HfCl_6$:0.008$Re^{4+}$ - m).

To gain deeper insight into the spectroscopic properties of $Re^{4+}$ ions, the Tanabe Sugano diagram can be analyzed (Figure 2a). Under a strong crystal field, the excited states of $Re^{4+}$ ions ($5d^3$ electronic configuration) undergo splitting due to spin-orbit coupling, generating several sublevels, *i.e.*, $^2T_{2g}(\Gamma_8)$, $^2T_{2g}(\Gamma_7)$, $^2T_{1g}(\Gamma_6)$ and $^2T_{1g}(\Gamma_8)$ (Figure 2a)[46,47]. Thus, upon ultraviolet light irradiation, electrons at $^4A_{2g}$ ground sate are excited to the $^4T_{2g}$ level, followed

by nonradiative relaxation to the $^{2}T_{2g}(\Gamma_7)$ excited level. After that, the radiation transition of $^{2}T_{2g}(\Gamma_7) \rightarrow {}^{4}A_{2g}$ transition takes place, leading to the observed emission around 730 nm. The analysis of the emission and excitation spectra of the representative $Cs_2HfCl_6$:0.008$Re^{4+}$ double perovskites indicates that when monitored at $\lambda_{em}$ = 728 nm excitation spectrum of this phosphor comprises an broad band in the range of 280-450 nm, which are assigned to the overlap between charge transfer (CT) transition of $[ReCl_6]^{2-}$ octahedron and the spin-allowed $^{4}A_{2g} \rightarrow {}^{4}T_{2g}$ transition of $Re^{4+}$ ions (Figure 2b)[43,46]. Upon the excitation at the optimal wavelength of 304 nm, the emission spectrum shown in Figure 2b) is dominated by a spectrally narrow emission band at 728 nm arising from the $^{2}T_{2g}(\Gamma_7) \rightarrow {}^{4}A_{2g}$ transition of $Re^{4+}$. In order to analyze the influence of dopant concentration on the spectroscopic properties of $Cs_2HfCl_6:Re^{4+}$ and to determine the optimal doping content of $Re^{4+}$ in $Cs_2HfCl_6$ lattice, the emission spectra of the $Cs_2HfCl_6:Re^{4+}$ with different concentration of dopant ions were measured (Figure 2c). It can be clearly noticed that at 93 K the shape of the emission band of $Re^{4+}$ remains not affected by the dopant concentrations (Figure 2c). However, the integral emission intensity increases gradually with enlargement of $Re^{4+}$ content up to $x = 0.008$ for which the maximum intensity value was reached (see Figure S4a)), whereas the concentration quenching effect occurs with further rising the dopant content. According to the Dexter theory, the involved concentration quenching mechanism results from electric dipole-dipole interaction (Figure S4b), with a critical distance of 40.67 Å (see Supporting Information for details).

The measurement of the luminescence decay curves of the $Cs_2HfCl_6:Re^{4+}$ ($\lambda_{exc}$=290 nm, $\lambda_{em}$=728 nm) recorded at 93 K confirm that the single exponential decay was observed in the case of all dopant concentrations and the $\tau$ shortens only slightly with increase of dopant

concentration from 0.05 ms for 0.002 $Re^{4+}$ to 0.04 ms for 0.050 $Re^{4+}$ (Figure 2e and f, Table S2). Furthermore, the shortening of the $^2T_{2g}(\Gamma_7)$ lifetime at high $Re^{4+}$ content further confirms the existence of concentration quenching in the designed double perovskites. Comparison of the excitation spectrum of the $Cs_2HfCl_6$:0.008$Re^{4+}$ measured at 93 K and 293 K (Figure 2 g and S5a) indicate that the excitation bands at 93 K become well-resolved within the 280-400 nm region, whereas these features become broadened and partially overlapped at 293 K. This phenomenon can be associated with enhanced phonon populations and electron-phonon interactions at elevated temperature. Moreover, the comparison of emission spectra of the $Cs_2HfCl_6$:0.008$Re^{4+}$ double perovskite recorded at 93 and 293 K presented in Figure S5b) further clarify the vibrational transition details of the sub-excited levels. At 93 K, highly resolved zero-phonon line (ZPL) together with multiple phonon sidebands (*i.e.*, $v_3$, $v_4$ and $v_6$), associated with vibrational modes of the octahedral units, are clearly distinguished (Figure 2c), which are independent of the doping content. Note that, as the temperature increases to 293 K, not only the emission intensity is altered, *i.e.*, the intensities of the anti-Stokes phonon sidebands are promoted, while those of the Stokes phonon sidebands are declined, but also the emission band is broadened (Figure S5b). This temperature-induced spectral evolution can be attributed to the increased phonon population and enhanced electron-phonon scattering, which broaden the individual vibronic transitions and redistribute the emission intensity among the phonon-assisted channels. These results provide compelling spectroscopic evidence for strong coupling between the $Re^{4+}$ electronic states and the vibrational modes of the surrounding $[ReCl_6]^{2-}$ octahedron.

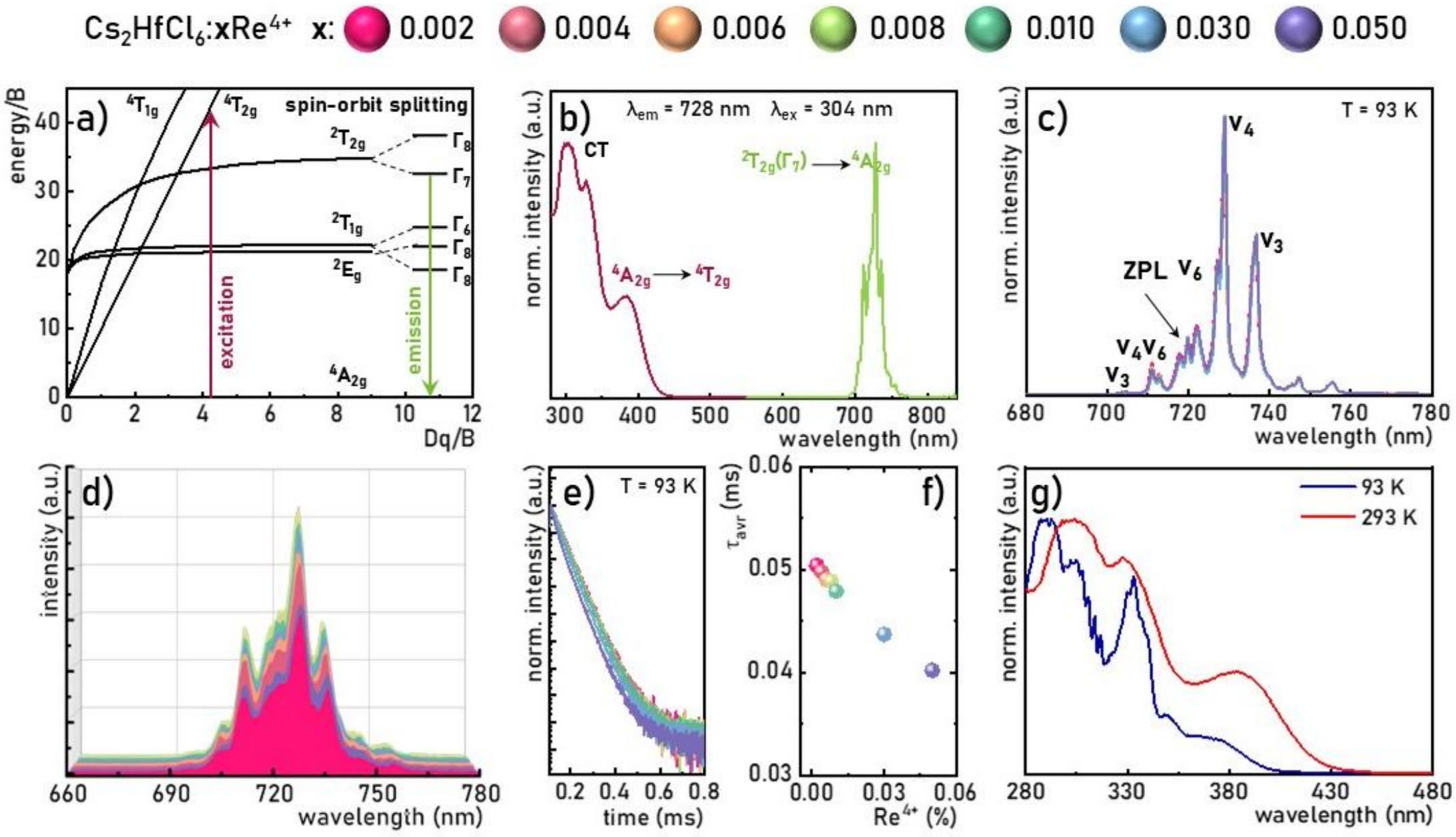

**Figure 2.** Tanabe-Sugano energy-level diagram of $Re^{4+}$ ions - a); excitation and emission spectra of the $Cs_2HfCl_6$:0.008$Re^{4+}$- b); comparison of the normalized measured at 93 K - c) and non-normalized measured at 303 K- d) emission spectra of the $Cs_2HfCl_6$:$Re^{4+}$ with different ions concentration; luminescence decay profiles - e) and $\tau_{avr}$ - f) of the $Cs_2HfCl_6$:$Re^{4+}$ with different ions concentration comparison of excitation spectra of $Cs_2HfCl_6$:0.008$Re^{4+}$ measured at 93K and 293 K - g).

In designing an ideal material for high-pressure sensing, one of the most critical parameters to consider is the mechanical stability of the host material under high-pressure conditions. To theoretically elucidate the structural evolution and stability of the $Cs_2HfCl_6$ host upon compression, the ab-initio calculation was conducted over the pressure range from ambient to 7 GPa. As demonstrated in Figure S7a), the lattice parameters (*i.e.*, $a = b = c$) decline gradually with increasing pressure, manifesting an approximately isotropic lattice contraction. As a result the unit cell volume continuously decreases as pressure arises (Figure S7b), implying the appreciable compressibility of the $Cs_2HfCl_6$ lattice under high-pressure conditions. The

pressure-induced lattice shrinkage is primarily associated with the shortening of the Hf-Cl bonds, which results in lattice densification. The electronic structure was further investigated to understand the effect of compression on the electronic properties of the $Cs_2HfCl_6$ lattice. At ambient pressure ($p$ = 0.00 GPa), the $Cs_2HfCl_6$ host pertains to a direct-type semiconductor characteristic with the band gap of 4.22 eV, where both the conduction band maximum and the valence band minimum all situate at Γ point (Figure 3a). It is worth to notice that, although the direct-band-gap character is maintained during compression, the band gap gradually decreases with pressure, *i.e.*, altering from 4.22 to 3.91 eV as the pressure increases from 0.00 to 7.00 GPa (Figure S8), indicating a pressure-induced narrowing of the electronic structure of the $Cs_2HfCl_6$ host. Furthermore, for the sake of assessing the lattice stability upon compression, the phonon dispersion calculations were performed along the high-symmetry directions of Brillouin zone at 0.00 and 7.00 GPa, as presented in Figure 3b) and 3c), respectively. Evidently, these calculated bands are relatively flat, and no imaginary frequencies are observed at low frequencies, suggesting that the $Cs_2HfCl_6$ host is dynamically stable over the investigated pressure range.

To experimentally verify the pressure-induced electronic structure evolution, the absorption spectra of the $Cs_2HfCl_6$ host as a function of pressure were examined and shown in Figure 3d). Evidently, an intense broad band arising from the $Cl^-$-$Hf^{4+}$ charge transfer within the isolated $[HfCl_6]^{2-}$ octahedron is gained in the recorded absorption spectrum. Moreover, this band is red-shift at elevated pressure because of the compression of the octahedra, demonstrating pressure-induced modulation of the electronic band structure. Moreover, when pressure releases, the absorption band returns its starting position (Figure 3d), implying the

good reversibility of the pressure-caused band gap regulation. Via using the following function, the pressure-dependent band gap ($E_g$) of the $Cs_2HfCl_6$ host was determined, as follows[43,45]:

$$\alpha h v = A\left(h v - E_g\right)^n \quad (1)$$

where the absorption coefficient is labeled by $\alpha$, $hv$ is the photon energy, $A$ is fitting constant and $n$ value is determined by the semiconductor type. Here, $n$ value is taken as 1/2 since $Cs_2HfCl_6$ pertains to a direct band-gap feature. Representative plots of the $(\alpha hv)^2$ $vs.$ $hv$ for the $Cs_2HfCl_6$ host recorded at 0.00, 4.35 and 8.18 GPa are illustrated in Figure S9a)-S9c), respectively. The band gap of the $Cs_2HfCl_6$ host declines gradually withe increasing pressure, which matches well with theoretical calculation results (Figure S8). These results confirm that compression provides an effective strategy for continuously regulating the electronic structure of the $Cs_2HfCl_6$ host.

Further experimental confirmation of the host material stability under high pressure conditions was done using *in-situ* high-pressure Raman spectra measurements Figure 3e). As demonstrated, with increasing the pressure (*i.e.*, 0.00-8.18 GPa), the Raman peaks gradually shift to higher wavenumbers, while no additional Raman peaks associated with impurity phases or structural transitions are observed, revealing that the resulting double perovskites possess good structure stability within the investigated pressure range. The shift of the Raman bands can be assigned to the pressure-induced the shortening of the bond length, which strengths the lattice vibrational frequencies. Moreover, the Raman peaks can revert to their initial positions after pressure release (*i.e.*, decompression process) (Figure S10). As it was shown the pressure dependence of the Raman band positions exhibits an approximately linear relationship, with the shift rates of 5.83 and 5.25 $cm^{-1}$ $GPa^{-1}$ for the $T_{2g}$ and $A_{1g}$ modes, respectively (Figure 4f and

4g). Apart from the linear shift of the Raman peaks, obvious peak broadening also observed at high-pressure (Figure 4e), which origins from the increased amount of crystal defects and strains in the contracted compounds. These results confirm that the $Re^{4+}$-doped $Cs_2HfCl_6$ double perovskites possess good structural robustness within the target pressure range, enabling their promising applications in high-pressure conditions.

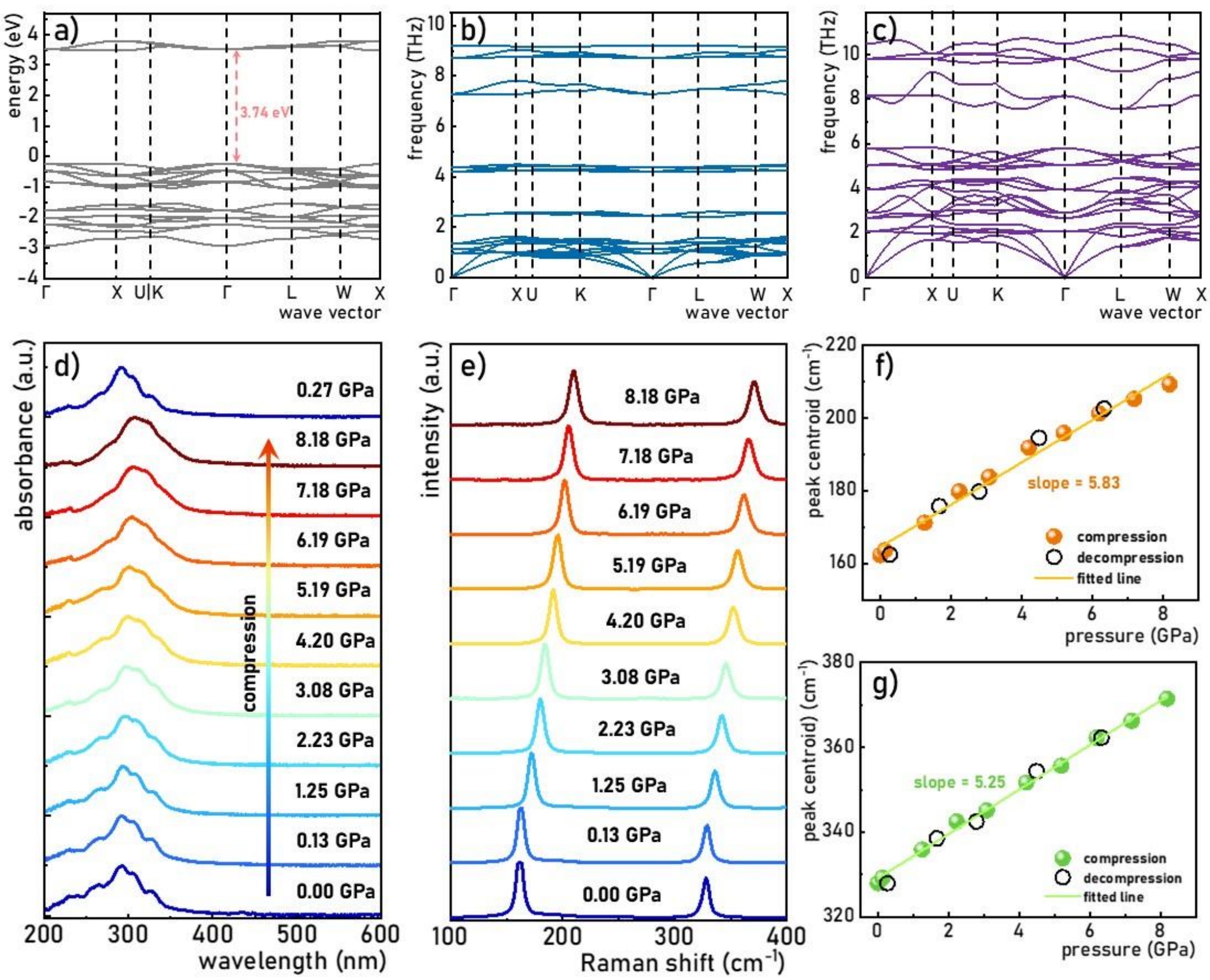


**Figure 3.** Calculated electronic band structure of the $Cs_2HfCl_6$ host at $p$ = 0.00 GPa; phonon dispersion spectra of the $Cs_2HfCl_6$ host at $p$ = 0.00 GPa- b) and $p$ = 7.00 GPa -c); absorption spectra of the $Cs_2HfCl_6$ host as a function of pressure -d) *in-situ* pressure dependent Raman spectra of the $CsHfCl_6$:0.008$Re^{4+}$-e) and the influence of the pressure on the Raman peak centroids of the $CsHfCl_6$:0.008$Re^{4+}$ - f) and -g).

Further assessment of the structural stability of the $Cs_2HfCl_6$:0.008$Re^{4+}$ at high pressure

conditions was performed based on the pressure-dependent XRD analysis (Figure 4a). Since the Mo K$\alpha$ irradiation ($\lambda$ = 0.7107 Å) was employed for the *in-situ* high-pressure XRD measurement, the positions of the collected diffraction peaks are inconsistent with those obtained by using Cu K$\alpha$ irradiation (Figure 1c). As disclosed, no additional diffraction peaks emerge and no original diffraction peaks vanish upon increasing pressure, indicating that the cubic phase of designed double perovskites remains stable throughout the investigated pressure range (0.13-9.03 GPa). Moreover, it is noted that the diffraction peaks not only monotonously move to the higher diffraction angles but also are broadened at high-pressure, revealing continuous lattice contraction and enhanced lattice distortion under compression. Here, the shift of the diffraction peaks results from the lattice contraction, while the broadened diffraction peaks are contributed by the pressure-triggered lattice strain and defect generation. Such compression induced structural changes are beneficial for regulating the crystal field environment surrounding $Re^{4+}$, thereby leading to pressure-dependent luminescence behaviors, as analyzed below. Note that, the diffraction peaks return to their original positions during decompression process (Figure 4a), confirming their good structural reversibility at high-pressure conditions. Furthermore, Rietveld XRD refinements were performed to get deeper insight into the structure evolution upon compression and the representative results are illustrated in Figure 4b) and 4c). Apparently, these refined diffraction profiles show excellent agreement with the experimental data, clarifying that the final products have pure cubic phase and the absence of pressure-induced phase transitions within the investigated pressure range. Although no macroscopic phase transition occurs during compression, the lattice parameters, *i.e.*, $a = b = c$ and cell volume, and Hf–Cl bond length decrease gradually as pressure increases,

as displayed in Table S3, Figure 4d) and 4e), indicating evident lattice contraction, which coincides well with the theoretical calculation results. In addition, with the aid of the third-order Birch-Murnaghan equation, we further analyzed the pressure-dependent cell volume, as defined below[45,48]:

$$p = \frac{3B_0}{2}\left[\left(\frac{V_0}{V}\right)^{7/3} - \left(\frac{V_0}{V}\right)^{5/3}\right] \times \left\{1 + \frac{3}{4}(B_0' - 4)\left[\left(\frac{V_0}{V}\right)^{2/3} - 1\right]\right\} \quad (2)$$

where $p$ refers to applied pressure, $V$ and $V_0$ denote the cell volume at recorded pressure $p$ and $p = 0$ GPa, respectively, $B_0$ stands for the bulk modulus at ambient condition and $B_0'$ represents the parameter for pressure derivate. Based on the fitting result, the $B_0$ value of the studied samples is determined to be 21.8 GPa, which is comparable with other halide double perovskites, such as $Cs_3Cu_2I_5$ ($B_0 = 29.6$ GPa), $(NH_4)_2SeBr_6$ ($B_0 = 22.28$ GPa), $Cs_2AgBiBr_6$ ($B_0 = 26.6$ GPa), *etc.*, [48–50] demonstrating the soft lattice characteristics of the $Re^{4+}$-doped $Cs_2HfCl_6$ double perovskites. Consequently, the use of external pressure provides an effective strategy to continuously regulate the local structural environment of $Re^{4+}$, enabling pressure-stimuli responsive luminescence.

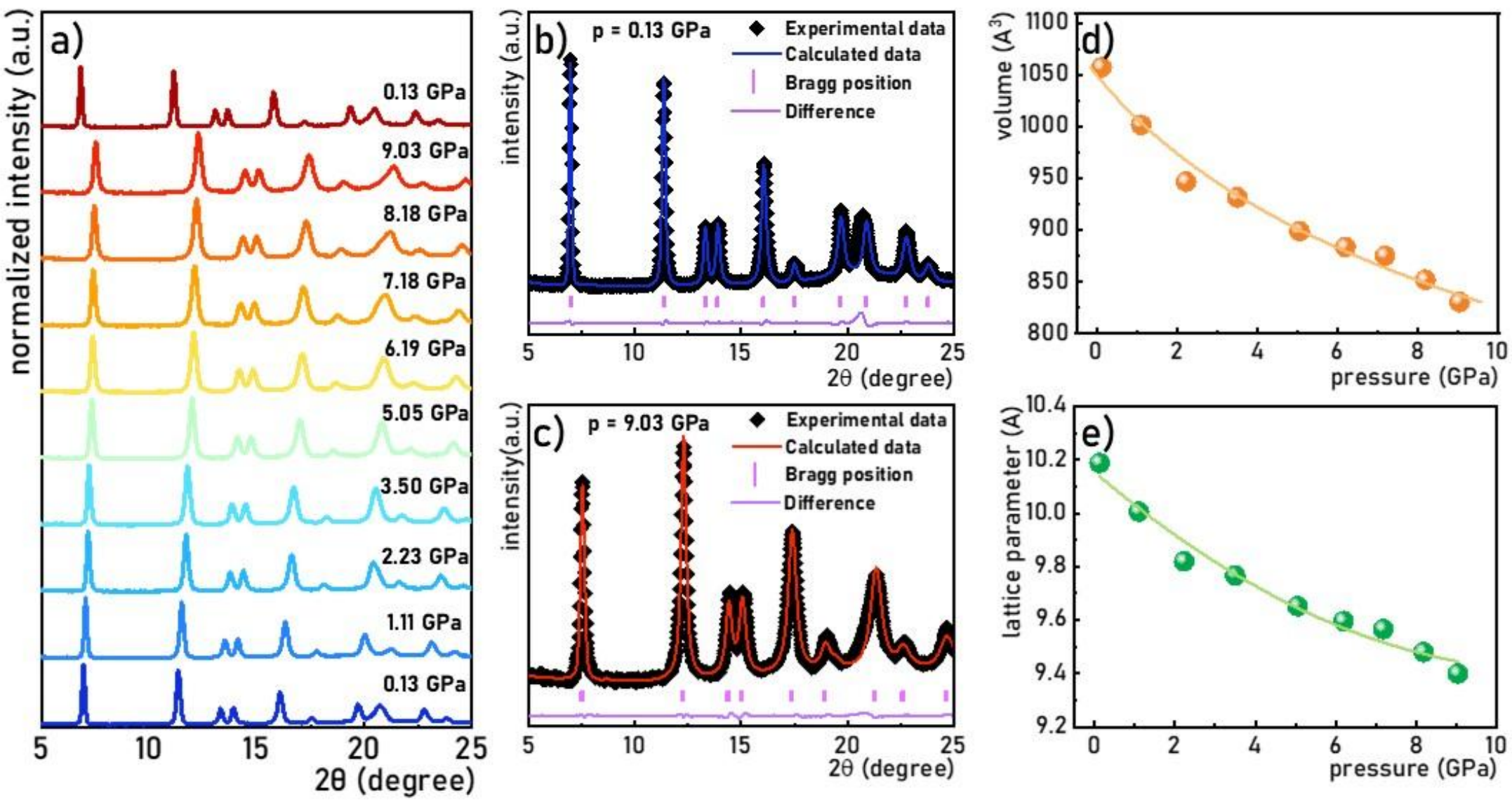


**Figure 4.** In-situ high-pressure XRD patterns of the $CsHfCl_6$:0.008$Re^{4+}$ - a); Rietveld refinements of the XRD

patterns recorded at the pressure of 0.13 GPa – b) and 9.03 GPa -c); The influence of the pressure on the unit cell volume -d) and lattice parameter (*i.e.*, $a = b = c$) -e) of the $CsHfCl_6$:0.008$Re^{4+}$.

One of the key advantages of transition-metal ions for luminescence manometry is the pronounced sensitivity of their spectroscopic properties to pressure[31]. This sensitivity primarily originates from two pressure-dependent effects: (i) changes in the crystal-field strength experienced by the transition-metal ions, reflected in variations in the energies of the $^4T_2$ and $^4T_1$ excited states, and (ii) changes in the nephelauxetic effect. Although the former generally induces substantially larger shifts in the spectral position of the emission bands, it may also introduce certain limitations for practical pressure sensing. As demonstrated for $Cr^{3+}$ ions in weak crystal fields, emission originating from the $^4T_2 \rightarrow ^4A_2$ electronic transition gives rise to a broad spectral band[51–55]. Such broadband emission can hinder precise pressure determination based on tracking the position of the emission maximum. Moreover, in high-pressure experiments performed, for example, using a diamond anvil cell, the broad spectral profile may significantly limit the applicability of such a material as a pressure indicator due to the increased risk of spectral overlap between the emission of the pressure calibrant and that of other luminescent materials investigated simultaneously.

An alternative approach exploits the pressure-induced modification of the nephelauxetic effect[11,13,16,56,57]. This mechanism is employed in one of the best-known and most widely used luminescent pressure calibrants, $Al_2O_3$:$Cr^{3+}$, commonly referred to as ruby[6,11–13,15]. In this material, $Cr^{3+}$ ions experience a strong crystal field and exhibit narrowband emission associated with the $^2E \rightarrow ^4A_2$ electronic transition. Application of pressure decreases the Cr-O bond length,

modifying the electronic interactions within the $CrO_6$ coordination environment and resulting in a redshift of the $Cr^{3+}$ emission lines. A major advantage of this approach is the narrow spectral width of the emission features, which enables highly precise pressure determination while simultaneously minimizing the risk of spectral overlap between the pressure calibrant and the luminescence of the material under investigation[6,11–13,15]. Therefore, the analogous narrowband luminescence of $Re^{4+}$ ions observed in $CsHfCl_6$:$Re^{4+}$, together with the similarity between the energy-level schemes of $Re^{4+}$ and $Cr^{3+}$, suggests that $CsHfCl_6$:$Re^{4+}$ may exhibit considerable potential as a luminescent material for optical pressure sensing. To verify this hypothesis the response of the luminescence characteristics of the target materials to pressure, *in-situ* pressure dependent emission spectra of $CsHfCl_6$:$Re^{4+}$ were collected using the experimental setup presented schematically in Figure 5a. The obtained results indicated a significant redshift of the $Re^{4+}$ emission band at the compression of the phosphor (Figure 6b). As disclosed, the characteristic featured emission of $Re^{4+}$ remains well preserved during compression. However, with elevating the pressure (*i.e.*, 0.00-6.93 GPa), a spectral redshift from 727.73 to 765.41 nm due to the lattice contraction induces a reduction in the Re-Cl bond length was observed. As a consequence, the emission wavelength gradually shifts toward longer wavelengths under compression[58,59]. Aside from the spectral red-shift, pressure-triggered luminescence intensity enhancement phenomenon in $CsHfCl_6$:$Re^{4+}$ was observed (Figure S11). Specifically, the emission intensity increases as pressure rises and achieves its maximum state at 4.64 GPa (*i.e.*, a 1.75-fold enhancement compared to the intensity at ambient pressure), and then it starts to decrease with further compression. Notably, the luminescence intensity does not show luminescence quenching below the value observed at ambient pressure even when the pressure

is elevated to 6.93 GPa (Figure 5c), manifesting the quenching-free luminescence behavior of the resultant double perovskites within the investigated pressure range. Furthermore, upon decompression, the emission band gradually shifts back toward shorter wavelengths (Figure 6d), which reveals the splendid reversibility of the pressure-caused luminescence evolution. These characteristics highlight the potential of the designed double perovskites for pressure sensing applications. The detailed analysis of the spectral position of the band position as a function of pressure reveals the monotonic change of the band maxima by ~37.68 which is over 15-fold larger than that of the ruby (*i.e.*, a displacement of 2.47 nm) (Figure 5e). The relation between pressure and emission band centroid satisfy a three-order polynomial expression, *i.e.*, $\lambda = -0.24p^2 + 6.93p + 728.55$. To quantify these changes the absolute manometric sensitivity can be calculated as follows:

$$S_{A(p)} = \frac{\Delta\lambda}{\Delta p} \tag{3}$$

The pressure-related absolute sensitivity of the $CsHfCl_6$:0.008$Re^{4+}$ decreases gradually at elevated pressure from 6.93 nm $GPa^{-1}$ at ambient pressure to 3.98 nm $GPa^{-1}$ at 6.9 GPa (Figure 5f). In the same pressure range the maximal absolute sensitivity obtained for ruby was equal to 0.365 nm $GPa^{-1}$ (Figure 5f). The comparison of the manometric performance of $CsHfCl_6$:$Re^{4+}$ with other previously reported narrow band emitting luminescence manometers reveals distinguished pressure sensing performance of the target materials (Figure 5g and Table 1). In addition, after two continuous compression-decompression cycles (Figure S12), it can be seen that the emission band centroid successfully revert its original state, revealing the excellent stability and reversibility of the studied samples. These results manifest that the $Re^{4+}$-doped $Cs_2HfCl_6$ double perovskites combine high pressure sensitivity with quenching-free

luminescence behavior, making them promising candidates for band-shift based optical manometry.

**Table 1** Comparison of narrow-band luminescence manometers exhibiting higher $S_A$ than ruby.

| ***Compound*** | ***Sensitivity (nm GPa$^{-1}$)*** | ***Pressure range (GPa)*** | ***Reference*** |
|---|---|---|---|
| ***$Al_2O_3$ (ruby)*** | **0.365** | 0-150 | [14] |
| ***$Al_2O_3$*** | **0.3796** | 0-2 | [27] |
| *LVMOF-1:$Eu^{3+}$* | 0.41 | 0.3-8 | [19] |
| *$MgO:Cr^{3+}$* | 0.504 | 0-7.9 | [20] |
| *$YPO_4:Yb^{3+}/Er^{3+}$* | 0.539 | 0-11.2 | [21] |
| *$YAlO_3:Cr^{3+}$* | 0.70 | 0-20 | [5] |
| *$SrGdAlO_4:Mn^{4+}$* | 0.79 | 0.03-7.6 | [22] |
| *$YPO_4:Yb^{3+}/Tm^{3+}$* | 0.80 | 0.64-24.35 | [21] |
| *$NaBiF_4:Yb^{3+}/Er^{3+}$* | 0.93 | 0.3-13.2 | [23] |
| *$\beta$-$Ga_2O_3:Cr^{3+}$* | 0.948 | 0-4.95 | [24] |
| *$BaFCl:Sm^{2+}$* | 1.10 | 0-20 | [5] |
| *$SrFCl:Sm^{2+}$* | 1.106 | 0-29 | [25] |
| *$Gd_2ZnTiO_6:Mn^{4+}$* | 1.11 | 0.03-8.01 | [26] |
| *$SrFCl:Sm^{2+}$* | 1.123 | 0-1.2 | [27] |
| *$Sr_4Al_{14}O_{25}:Mn^{4+}$* | 1.20 | 0.03-7.6 | [28] |
| *$GdTaO_4:Nd^{3+}$* | 1.34 | 0-8.65 | [29] |

| $YVO_4:Yb^{3+}/Er^{3+}$ | 1.766 | 0-6.5 | [30] |
|---|---|---|---|
| **$Cs_2HfCl_6:Re^{4+}$** | **6.93** | **0-6.93** | **This work** |

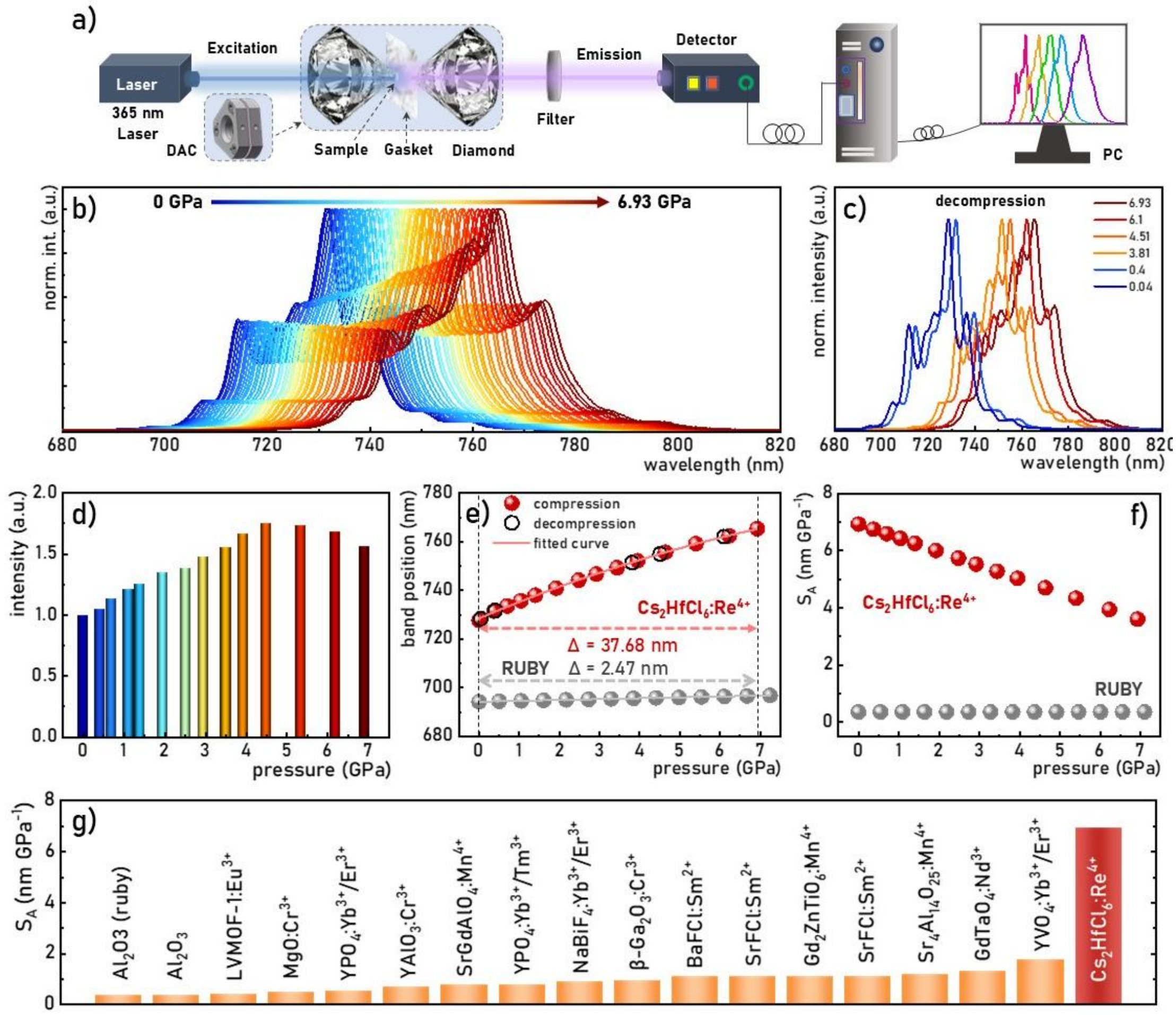


47

**Figure 5.** Schematic illustration of the high-pressure luminescence measurement setup used in the experiment – a); normalized emission spectra of $CsHfCl_6:0.008Re^{4+}$ measured as a function of pressure during compression – b) and decompression – c) the influence of the applied pressure on the integral emission intensity of $Re^{4+}$ ions in $CsHfCl_6:0.008Re^{4+}$ -d); comparison of the pressure dependence of the emission band centroid of $Re^{4+}$ ions in $CsHfCl_6:0.008Re^{4+}$ and $Cr^{3+}$ ions in ruby -e) and corresponding pressure dependence of $S_A$ -f); comparison of $S_A$ for different luminescent manometers based on narrow band emission – g).

An essential aspect in the development of luminescent pressure sensors is the verification of the specificity of their spectroscopic response to pressure. In this context, it is crucial to determine how the investigated spectroscopic parameters are influenced by variations in another equally important thermodynamic variable, namely temperature. Therefore, spectroscopic properties of the $Cs_2HfCl_6$:0.008$Re^{4+}$ were analyzed in the 93-433 K thermal range. As disclosed, the luminescence intensity shows a monotonous downward tendency with temperature, which results from the thermal quenching effect (Figure 6a). Note that, although the synthesized double perovskites suffer from thermal quenching effect, its emission band position is only slightly thermally shifted (*i.e.*, a faint blue-shift), indicating excellent spectral stability and favorable characteristics for optical sensing applications (Figure 6b). Furthermore, through analyzing the relation between temperature and luminescence intensity, the activation energy ($\Delta E$) governing the thermal quenching is estimated, as described below[47]:

$$I = \frac{I_0}{1 + A\exp(-\frac{\Delta E}{k_B T})} \tag{4}$$

where I and $I_0$ refers to the emission intensities at measured and initial temperature, respectively, $A$ is the fitting constant and $k$ is assigned to the Boltzmann constant. Based on the linear fitting of $\ln(I_0/I\text{-}1)$ *vs.* $1/kT$, one achieves that the activation energy for the $Re^{4+}$ in $Cs_2HfCl_6$ lattice is 0.04 eV. Based on the emission spectra of $Cs_2HfCl_6:Re^{4+}$ recorded as a function of temperature, the temperature dependence of the spectral position of the $Re^{4+}$ emission band can be determined (Figure 6c). The obtained results reveal that, over the investigated temperature range, increasing temperature induces a systematic blueshift of the emission band from 728.9

nm at 93 K to 726.6 nm at 435 K. These data enable the absolute thermal sensitivity ($S_A$) to be determined according to the following equation:

$$S_{A(T)} = \frac{\Delta\lambda}{\Delta T} \quad (5)$$

For $Re^{4+}$, the $S_A$ value increases monotonically from 0.0042 nm $K^{-1}$ at 113 K to 0.0098 nm $K^{-1}$ at 435 K (Figure 6d). Importantly, the obtained $S_A$ values are only slightly higher than those determined for ruby, for which $S_A$ increases from 0.00381 nm $K^{-1}$ at 113 K to 0.009 nm $K^{-1}$ at 453 K. The relatively small difference in the thermal $S_A$ of these two materials is particularly noteworthy considering the substantially higher absolute manometric sensitivity exhibited by $Cs_2HfCl_6:Re^{4+}$. This combination of a strongly enhanced pressure response with only a minor increase in temperature-induced spectral changes highlights the favorable characteristics of $Cs_2HfCl_6:Re^{4+}$ for luminescence-based pressure sensing with reduced temperature interference.

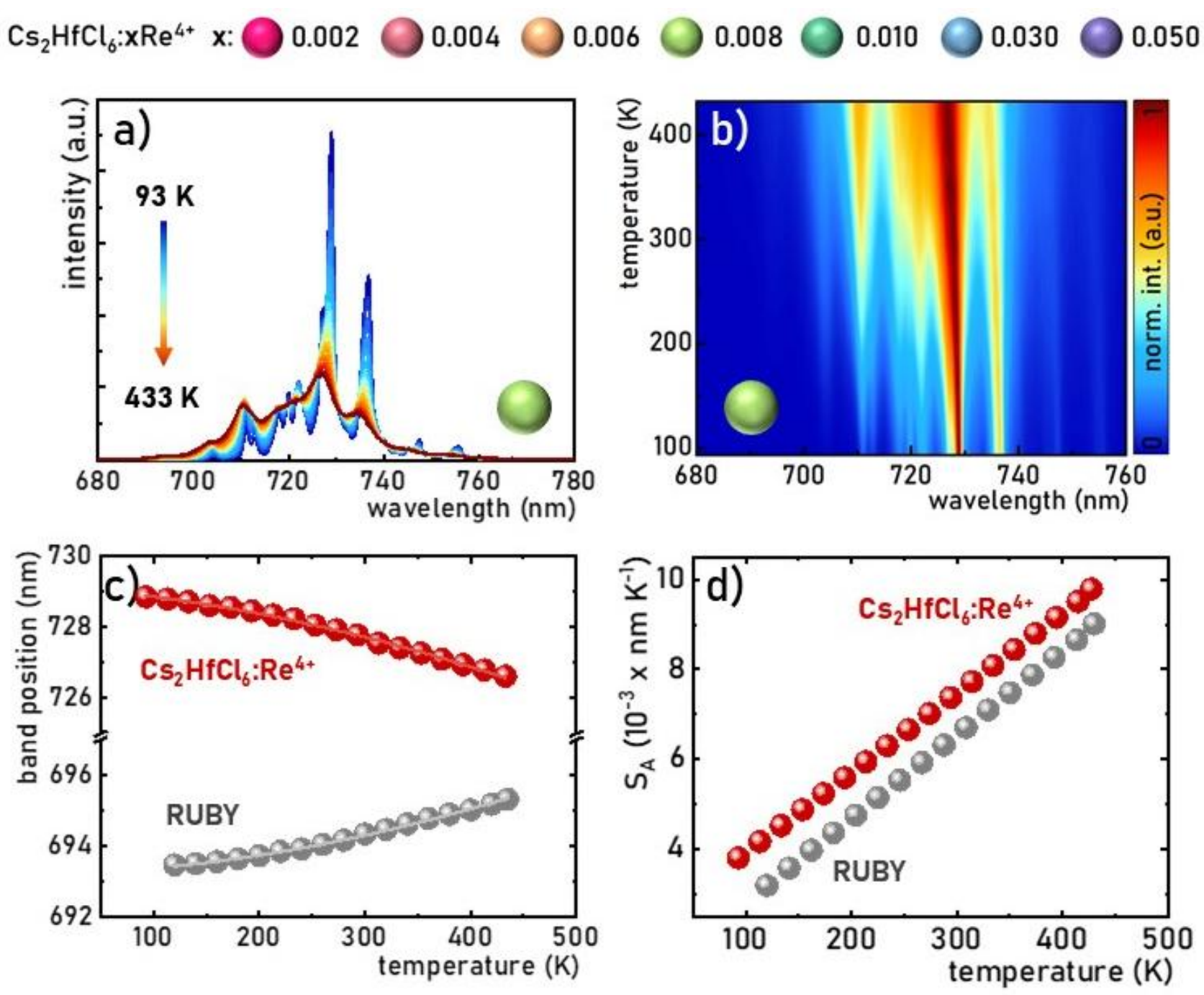


**Figure 6.** Temperature-dependent emission spectra of $CsHfCl_6:0.008Re^{4+}$ -a) and the corresponding

luminescence map of normalized spectra -b); thermal dependence of the band position of $CsHfCl_6:0.008Re^{4+}$ and ruby – c) and corresponding thermal dependence of $S_A$ -d).

Although, as demonstrated above, pressure determination based on the spectral shift of the emission band offers several important advantages, particularly for measurements performed in a DAC, the pressure-induced spectral shift of the $Re^{4+}$ emission can also be exploited to develop a ratiometric pressure-sensing approach. For this purpose, two appropriately selected spectral ranges can be defined on opposite sides of the emission band, such that the pressure-induced redshift results in opposite changes in their integrated emission intensities[52,54]. The ratio of these intensities can subsequently serve as a ratiometric parameter for pressure determination. Importantly, temperature variations are not expected to significantly modify the nephelauxetic effect and, consequently, should have only a minor influence on the spectral position of the $Re^{4+}$ emission band. Therefore, a ratiometric parameter based predominantly on the pressure-induced spectral shift is expected to exhibit relatively low cross-sensitivity to temperature. This feature is particularly advantageous for reliable pressure sensing under conditions where both pressure and temperature may vary simultaneously.

Herein, the emission band was integrated over two fixed wavelength intervals with the bandwidth of 10 nm, *i.e.*, I1 = 720-730 nm and I2 = 770-780 nm, as presented in Figure 7a). The pressure induced redshift of the emission band of $Re^{4+}$ ions results in a monotonic over 100-fold enhancement of the emission signal integrated into I2 in the analyzed pressure range. On the other hand in the same pressure range the analogous quenching of the emission intensity integrated within I1 was observed. This opposite manometric response of the luminescent

signals in I1 and I2 enables to propose the following luminescence intensity ration (*LIR*) as a manometric parameter:

$$LIR = \frac{\int_{720nm}^{730nm} I(\mathrm{Re}^{4+}: {}^2T_{2g}(\Gamma 7) \rightarrow {}^4T_{2g}) d\lambda}{\int_{770nm}^{780nm} I(\mathrm{Re}^{4+}: {}^2T_{2g}(\Gamma 7) \rightarrow {}^4T_{2g}) d\lambda} \quad (6)$$

As can be seen in the analyzed pressure range *LIR* preserves monotonic enhancement (Figure 7c). Furthermore, for the sake of quantitatively clarifying the sensing performance of the ratiometric approach, we calculated the relative pressure sensitivity ($S_{R,p}$), as defined below:

$$S_{R(p)} = \frac{1}{LIR}\frac{\Delta LIR}{\Delta p}100\% \quad (7)$$

As can be clearly seen, the $S_{R,p}$ value promotes as pressure elevates, achieving its maximum value of $S_{R,p,Max}$ = 175.1% GPa$^{-1}$ at around 3.84 GPa. Furthermore, the $S_{R,p}$ value remains higher than 22.0% GPa$^{-1}$ over the whole operating pressure range, implying the excellent pressure detection capability of the proposed ratiometric strategy. It is crucial to underline that *LIR* does not reveal significant thermal change in the 93-453K thermal range which result in a very small thermal relative sensitivity which does not exceed 0.16 % K$^{-1}$ and reaches $S_{R,T}$=0.11% K$^{-1}$ at room temperature (Figure 7e). To evaluate the selectivity of a luminescent manometer toward pressure variations and quantify its susceptibility to temperature-induced cross-sensitivity, the Thermal Invariability Manometric Factor (TIMF) can be determined according to the following expression[60]:

$$TIMF = \frac{S_{R,p}}{S_{R,T}} \quad (8)$$

where $S_{R,p}$ and $S_{R,T}$ represent the relative sensitivities to pressure and temperature, respectively. Since measurements of the pressure dependence of *LIR* at different temperatures are often

experimentally challenging, a simplified approach is commonly employed in which a single, constant value of the thermal sensitivity, typically determined at room temperature, is used for *TIMF* calculations. A similar approach was adopted in the present study.

The obtained results clearly demonstrate that the *TIMF* of $CsHfCl_6$:0.008$Re^{4+}$ increases with pressure, reaching a value of approximately 1590 K $GPa^{-1}$above 2 GPa. This value remains nearly constant upon further compression up to approximately 5 GPa, above which a gradual decrease in *TIMF* is observed. Nevertheless, the obtained *TIMF* value is remarkably high. Physically, it represents the magnitude of the temperature variation required to induce a change in LIR equivalent to that produced by a pressure variation of 1 GPa. Therefore, the high TIMF obtained for $CsHfCl_6$:0.008$Re^{4+}$ confirms the pronounced selectivity of its ratiometric luminescence response toward pressure relative to temperature.

A comparison of the manometric relative sensitivity of $CsHfCl_6$:0.008$Re^{4+}$ with those of previously reported ratiometric luminescent manometers further highlights its excellent performance. The relative pressure sensitivity achieved for $CsHfCl_6$:0.008$Re^{4+}$ is the second-highest value reported in the literature to date. Among the luminescent manometers considered, only $La_6Sr_4(SiO_4)_6F_2$:$Ce^{3+}$ exhibits a higher relative pressure sensitivity[61]. These results demonstrate the considerable potential of $CsHfCl_6$:0.008$Re^{4+}$ as a highly sensitive ratiometric luminescent manometer with a favorable resistance to temperature-induced interference. Obtained results clearly indicate that the $Re^{4+}$-doped $Cs_2HfCl_6$ double perovskites show great potential for highly-sensitive ratiometric optical manometers.

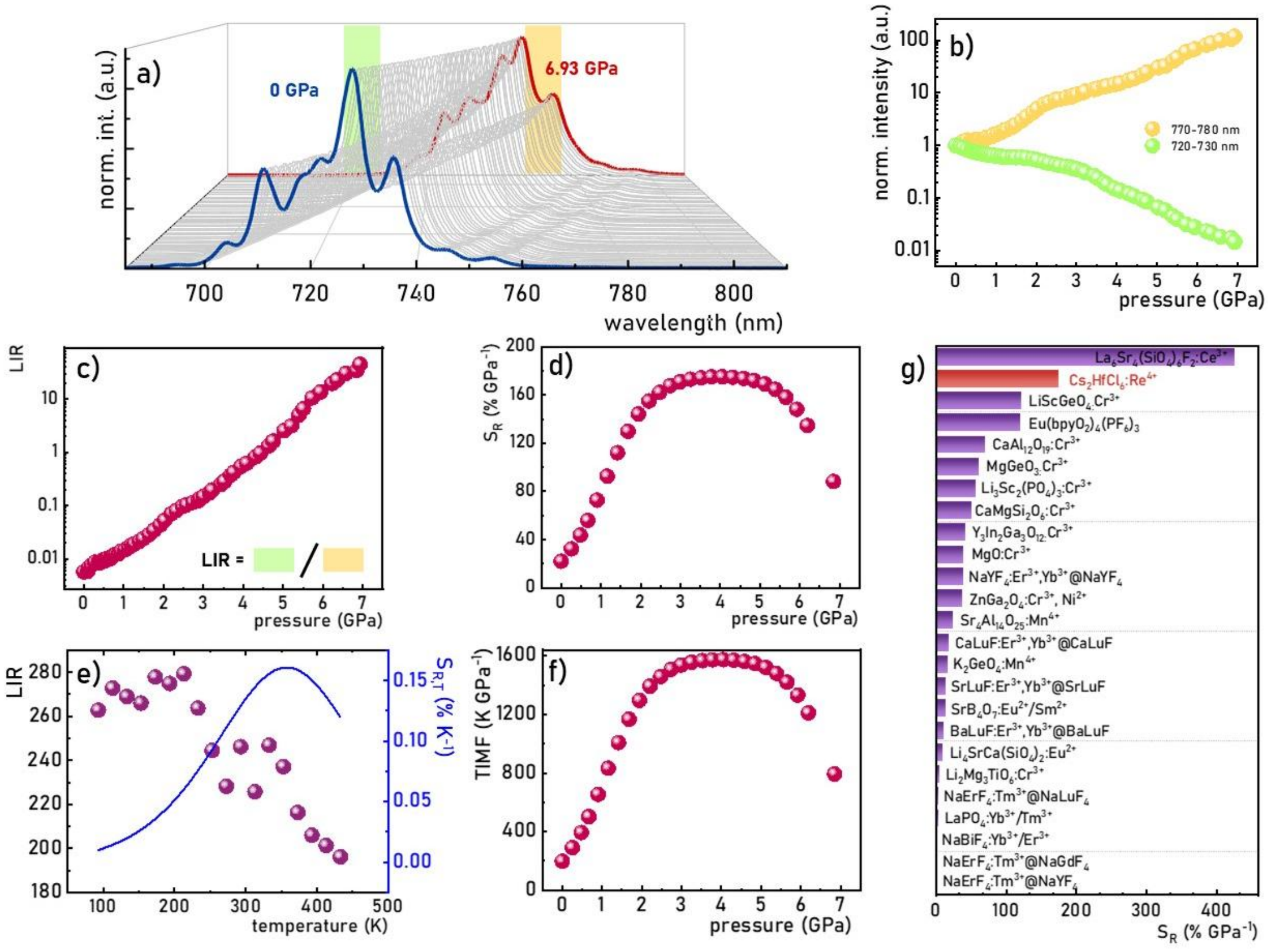

**Figure 7.** Schematic representation of the pressure dependent emission spectra of $CsHfCl_6:0.008Re^{4+}$ with marked two spectral ranges used in ratiometric approach -a); the influence of pressure on the emission intensity of $Re^{4+}$ ions integrated within two different spectral ranges -b) pressure dependence of *LIR*-c) and corresponding manometric $S_R$- d); thermal dependence of *LIR* and $S_{RT}$-e) and pressure dependence of *TIMF* -f) comparison of the $S_R$ for different ratiometric luminescence manometers -g)[21,23,51,52,55,61–65].

## Discussion

In this work, the application potential of $CsHfCl_6:0.008Re^{4+}$ for remote pressure sensing was investigated as an alternative to ruby, which is widely regarded as the gold standard for luminescence-based pressure sensing. For this purpose, the spectroscopic properties of $CsHfCl_6:0.008Re^{4+}$ were systematically investigated as a function of both temperature and pressure. The analysis revealed that $CsHfCl_6:0.008Re^{4+}$ exhibits intense, narrowband

luminescence centered at approximately 728 nm, associated with the $^2T_{2g}(\Gamma_7) \rightarrow {}^4A_{2g}$ electronic transition of $Re^{4+}$ ions. Increasing pressure, accompanied by shortening of the Re-Cl bond, induces a gradual redshift of the $Re^{4+}$ emission band at a rate of 6.93 nm $GPa^{-1}$, which is over 15 times higher than that observed for ruby. To the best of our knowledge, this represents the highest absolute pressure sensitivity reported to date for a luminescent manometer based on narrowband emission.

An important advantage of $CsHfCl_6:0.008Re^{4+}$ is its relatively low sensitivity to temperature variations. The spectral position of the $Re^{4+}$ emission band exhibits only a minor temperature-induced shift, with a rate comparable to that observed for ruby. Consequently, the substantially enhanced pressure response of $CsHfCl_6:0.008Re^{4+}$ is achieved without a corresponding increase in its susceptibility to thermal interference, which is particularly advantageous for reliable pressure sensing.

Furthermore, $CsHfCl_6:0.008Re^{4+}$ was demonstrated to enable ratiometric pressure readout based on the ratio of luminescence intensities integrated over two selected spectral ranges. The relative pressure sensitivity of the resulting ratiometric parameter reaches 175.1% $GPa^{-1}$, placing $CsHfCl_6:0.008Re^{4+}$ among the most sensitive ratiometric luminescent manometers reported to date. Overall, the combination of intense narrowband emission, exceptionally high pressure sensitivity, limited temperature dependence, and the possibility of ratiometric pressure readout demonstrates that $CsHfCl_6:0.008Re^{4+}$ represents a promising alternative to ruby as a luminescent pressure indicator, particularly in the pressure range below 7 GPa.

## Materials and methods

*Synthesis of the $Re^{4+}$-doped $Cs_2HfCl_6$ double perovskites*

Utilizing a hydrothermal method, a series of $Cs_2HfCl_6$:$x$$Re^{4+}$ ($0.002 \leq x \leq 0.050$) double perovskites were synthesized. Here, the powders of CsCl, $HfCl_4$ and $K_2ReCl_6$ were used as the raw materials. According to the stoichiometric ratio, these above raw materials were weighed and dissolved in 5 mL of HCl (37%). Then, the mixture was transferred into an autoclave and kept in an oven with the fixed temperature of 200 ºC for 12 h. When the temperature naturally cooled, the final products were washed with HCl and ethanol for three times. Ultimately, the dry process was performed and the designed double perovskites were collected for further characterizations.

*Sample characterization*

With the help of an X-ray diffractometer (Bruker D8 Advance; Cu K$\alpha$ radiation $\lambda = 1.5406$ Å), the phase structure of the prepared samples was examined. The elements presented in the synthesized double perovskite were explored through an X-ray photoelectron spectrometer (PHI 5000 VersaProbe III). Via the use of a field-emission scanning electron microscope (FE-SEM; Hitachi SU3500) and a transmission electron microscope (TEM; JEOL JEM-2100F), the morphological features of the studied samples were measured. The emission and excitation spectra of designed compounds at ambient conditions were tested by a fluorescence spectrometer (Edinburgh FS5). The temperature-dependent decay curves were measured using a Fluorescence spectrometer (Edinburgh FLS1000), which was equipped with a 150 W μFlash lamp and R928 photomultiplier tube from Hamamatsu. Furthermore, the temperature of the sample was controlled by using a heating-cooling stage (Linkam THMS600). Additionally, the calculated details were summarized in the Supporting Information.

Through employing a symmetric DAC, the high-pressure measurements were conducted, of which the generated pressure DAC was calibrated by monitoring the fluorescence of ruby. As for the sample chamber, it was a T301 steel gasket, whose thickness was 40 μm and a hole (*i.e.*, 150 μm diameter) was drilled. The resulting double perovskites were loaded into the cavity and the silicone oil was used as the pressure-transmitting medium. Using a 365 nm laser as excitation lighting source, the pressure-dependent emission spectra were collected by means of a microscopic fluorescence measurement system (MicroHP-SP). Furthermore, in-situ high-pressure Raman spectra were measured through a Mono Vista CRS+500 micro-Raman spectrometer and the excitation source was a 532 nm laser. Furthermore, the high-pressure absorption spectra of the resulting double perovskites were tested by an Ocean Optics Pro spectrometer. Ultimately, in-situ high-pressure XRD patterns were checked by means of a Rigaku Nanopix We diffactometer (Mo K$\alpha$ radiation $\lambda$ = 0.7107 Å).

**Acknowledgements**

This work was supported by the Foundation for Polish Science under First Team FENG.02.02-IP.05-0018/23 project with funds from the 2nd Priority of the Program European Funds for Modern Economy 2021-2027 (FENG). This work was supported by National Natural Science Foundation of China (62575147) and Ningbo Youth Science and Technology Innovation Leading Talent Project (2025QL025).